\documentclass[showpacs,preprintnumbers,amssymb]{revtex4}
\usepackage{epsfig}

\usepackage{threeparttable}

\def\beq{\begin{eqnarray}}
\def\eeq{\end{eqnarray}}
\def\nnb{\nonumber}

\newcommand{\be}{\begin{equation}}
\newcommand{\ee}{\end{equation}}
\newcommand{\bea}{\begin{eqnarray}}
\newcommand{\eea}{\end{eqnarray}}
\newcommand{\ba}{\begin{array}}
\newcommand{\ea}{\end{array}}
\begin{document}
\title{Lepton Flavor Violating $\tau$ Decays
in Two Higgs Doublet Model III}
\author{Wen-Jun Li$^{1,2}$}
\email{liwj24@163.com}
\author{Ying-Ying Fan$^{1}$, Gong-Wei Liu$^{1}$, Lin-Xia L\"{u}$^{3}$}
 \affiliation{$^1$ Department of Physics, Henan
Normal University, XinXiang, Henan, 453007, P.R.China \nnb \\
$^2$ Institute for Theoretical Physics China, CAS, Beijing, 100190,
P.R.China\nnb \\
$^3$ Department of Physics, Nanyang Normal University, Nanyang,
Henan, 473061, P.R.China}

\begin{abstract}

We study the lepton flavor violating processes of
$\tau^-\to\mu^-PP$ decays with
$PP=K^+K^-,K^0\bar{K}^0,\pi^+\pi^-,\pi^0\pi^0$ in the framework of
two Higgs doublet model III by virtue of the Chiral
Perturbation Theory. In this model, only three neutral
Higgs bosons contribute to these decays. With the current
 experimental constraints, we show that (a) the contributions of
 $|\lambda_{uu(dd)}|$ term are very small for these four decays;
 (b) we get the correlation between $|\lambda_{ss}|$ and
 $|\lambda_{\tau\mu}|$.
 For $|\lambda_{\tau\mu}| \sim 10-400$, one has $|\lambda_{ss}|\sim
 40-1$;
(c) in the existing model parameter space, $Br(\tau^- \to \mu^-
K^+K^-)$ could reach the order of ${\cal O}(10^{-8})$, but $Br(\tau
\to \mu^- \pi^+\pi^-/\pi^0\pi^0)$ are too small to be observed.
\end{abstract}


\maketitle

\noindent
\section{\bf Introduction}

    Flavor physics has made rapid development in these decades.
In addition to B physics, $\tau$ physics, including the
determination of $\alpha_s$ from the inclusive hadronic width, tests
of charged-current universality, and lepton flavor violation (LFV)
decays, etc., also belong to one branch of flavor physics. Among
these topics, LFV decays became more important after the discovery
of neutrino flavor oscillations and related non-zero neutrino
masses~\cite{superK1,K2,K3}. The LFV decays are forbidden or
suppressed strongly in the Standard Model (SM). While for new
physics, their decay ratios would  be enhanced largely by new LFV
sources and/or new particles effects. Obviously, LFV decays would be
a sharp tool to seek for the new physics beyond the SM. Due to its
large mass, $\tau$ lepton has many LFV decay modes, such as
$\tau\rightarrow 3l^\prime,~\tau\rightarrow l^\prime \gamma $ and
$\tau\rightarrow l^\prime M(M=P,V^0, PP)$, which could display rich
information on LFV mechanism. In particular, in certain new physics
models, e.g., where the LFV effects arise through the Higgs boson
exchanges, the $\tau$ lepton is expected to have large LFV couplings
to the $\mu$ because of its large mass. Therefore, $\tau$ LFV decays
become an interesting field for finding new physics.

  Among $\tau$ various LFV
decays, $\tau \to 3l',\tau \to \mu \gamma$ and $\tau \to l'P(V^0)$
have been searched in experiment~\cite{exp} and investigated in many
theoretical frameworks, such as realistic non-minimal supergravity
models~\cite{SUG}, SUSY SO(10) models~\cite{SO1,SO2}, various
extended-MSSM models ~\cite{MSSMNR,MSSM2}, Seesaw models
~\cite{saw1,saw2,saw3}, Unparticle scenario~\cite{unparticle},
Topcolor models~\cite{top}, the Littlest Higgs model with
T-Parity~\cite{lht}, etc. For $\tau^- \to \mu^- PP$ decays, the
newest experimental upper limits
 of $PP=K^+K^-(K^0\bar{K}^0,\pi^+\pi^-,\pi^0\pi^0)$
decays are ~\cite{exp,data1,data2}:
\bea &&{\cal B}(\tau^- \to \mu^-
K^+K^-)<6.8 \times 10^{-8},\,\,\,\,90\%CL~,~\,
\nnb \\
&&{\cal B}(\tau^- \to \mu^- \pi^+\pi^-)<3.3 \times
10^{-8},\,\,\,\,90\%CL ~,~\,\nnb
\\&&{\cal B}(\tau^- \to \mu^- K^0_sK^0_s)<8.0 \times
10^{-8}, \,\,\,\,90\%CL ~,~\, \nnb \\&&{\cal B}(\tau^- \to \mu^-
\pi^0\pi^0)<1.4 \times 10^{-5}, \,\,\,\,90\%CL ~,~\,\nnb
 \eea
  where the first three values have been improved by roughly one or two
 orders of magnitudes. Accordingly,
 their theoretical investigations have been
done in many possible extensions of the
SM~\cite{Ilakovac,chen,Herrero,Arganda,Yue}. C. Chen {\it et al.}
have analyzed the scalar bosons effects on $\tau^- \to \mu^-
K^+K^-(K^0\bar{K}^0)$ decays in the supersymmetric seesaw model with
nonholomorphic terms in the lepton sector at the condition of large
$\tan \beta$~\cite{chen}. E.~Arganda $\textit{et al.}$ have
investigated these processes in two constrained seesaw scenarios in
the Minimal Supersymmetric Standard Model (MSSM)~\cite{Arganda}.
M.~Herrero {\it et al.} have found that $\tau^-\to \mu^- \eta$ and
$\tau^- \to \mu^- f_0$ are more efficient to test indirectly to the
Higgs sector than $\tau \to 3\mu, \tau^- \to \mu^- K^+K^-$ decays in
supersymmetric seesaw models~\cite{Herrero}. And C. Yue $\textit{et
al.}$ has calculated the new particle effects in the
topcolor-assisted technicolor model and the littlest Higgs model
with T parity~\cite{Yue}.

In this paper, we shall study $\tau$ decays in the Two Higgs Doublet
Model III (2HDM III) since it naturally introduces flavor-changing
neutral currents (FCNCs) at tree level. In order to satisfy the
current experiment constraints, the tree-level FCNCs are suppressed
in low-energy experiments for the first two generation fermions.
While processes concerning with the third generation fermions would
be larger. Therefore, these FCNCs with neutral Higgs bosons mediated
may produce sizable effects on the $\tau - \mu$ transition.
Moreover, the evaluation of neutral Higgs effects would help to
provide useful information to identify the Higgs signals in LHC
experiment. Hitherto, many works have been performed in this
scenario. For example, the authors in~\cite{Matsuzaki} have
discussed the $\tau \to 3\mu$ decay in the 2HDM III and MSSM, and
analyzed their collider signals. $\tau^\pm \to l^\pm P^0$ decays
have been considered in the 2HDM III model with four-texture Yukawa
couplings~\cite{4Y}, which is shown that the model parameters in
leptonic sector are less constrained by present experimental data.
$\tau \to \mu P(V^0)$ decays have been studied in our previous
paper~\cite{li1,li2}, where the hadronization in final state is
merely expressed in terms of the meson decay constants and meson
masses. Furthermore, $\tau \to \mu$ transition mediated via Higgs
bosons is sensitive to the Higgs sector. In theory, the Higgs
influences on LFV decays have argued in 2HDM III and
MSSM~\cite{Shinya04,Shinya06,higgs1,higgs2}. Aoki {\it et
al.}~\cite{Aoki} have disputed the decay properties of Higgs bosons
in four types of 2HDMs and their collier phenomenology. So we extend
our discussion to the case of two pseudoscalar mesons in the
hadronic final state and deal with hadron matrix elements by the
chiral perturbative Theory~$(\chi P T)$~\cite{Pt1,Pt2,Pt3}.

The paper is organized as follows. In Section II, we make a brief
introduction of the theoretical framework for the 2HDM III.
In Section III, we deliberate the calculations of the
decay amplitudes with $\chi P T$ and our numerical predictions. Our
conclusions are given in the last section.

\section{\bf The Two-Higgs -Doublet Model III}

As the simplest extension of the SM, the Two-Higgs-Doublet Model has
an additional Higgs doublet. In order to ensure the forbidden FCNCs
at tree level, it requires that either the same doublet couple to
the \textit{u}-type and \textit{d}-type quarks (2HDM I) or one
scalar doublet couple to the \textit{u}-type quarks and the other to
\textit{d}-type quarks (2HDM II). While in the 2HDM
III~\cite{cheng,2hdm31,2hdm32,2hdm33,2hdm34,2hdm35}, the two Higgs
doublets could couple to the \textit{u}-type and \textit{d}-type
quarks simultaneously. Particularly, without an {\it ad hoc}
discrete symmetry exerted, this model does have FCNCs at the tree
level.

The Yukawa Lagrangian is generally expressed as the following form
\beq {\cal L}_{Y}= \eta^{U}_{ij} \bar Q_{i,L} \tilde H_1 U_{j,R} +
\eta^D_{ij} \bar Q_{i,L} H_1 D_{j,R} + \xi^{U}_{ij} \bar
Q_{i,L}\tilde H_2 U_{j,R} +\xi^D_{ij}\bar Q_{i,L} H_2 D_{j,R} \,+\,
h.c., \label{lyukmod3}
 \eeq
 where $H_i(i=1,2)$ are the two Higgs doublets.
 $Q_{i,L}$ is the left-handed
 fermion doublet, $U_{j,R}$ and
$D_{j,R}$ are the right-handed singlets, respectively. These
$Q_{i,L}, U_{j,R}$ and $D_{j,R}$ are weak eigenstates, which can be
rotated into mass eigenstates. While $\eta^{U,D}$ and $\xi^{U,D}$
are the non-diagonal matrices of the Yukawa couplings.

We can conveniently choose a suitable basis to denote $H_1$ and
$H_2$ as
 \bea
 \label{base}
 H_1=\frac{1}{\sqrt{2}}\left[\left(\ba{c} 0 \\
v+\phi^0_1 \ea\right)+ \left(\ba{c} \sqrt{2}\, G^+\\
i G^0\ea\right)\right], \,\,\,\,\,\,\,\,
 H_2=\frac{1}{\sqrt{2}}
 \left(\ba{c}\sqrt{2}\,H^+\\ \phi^0_2+i A^0\ea\right),
 \eea
where $G^{0,\pm}$ are the Goldstone bosons, $H^{\pm}$ and $A^0$ are
the physical charged-Higgs boson and CP-odd neutral Higgs boson,
respectively. Its virtue is that the first doublet $H_1$ corresponds
to the scalar doublet of the SM while the new Higgs fields arise
from the second doublet $H_2$.

The CP-even neutral Higgs boson mass eigenstates $H^0$ and $h^0$ are
linear combinations of $\phi_1^0$ and $\phi^0_2$ in Eq.~(\ref{base})
as follows
 \bea \label{masseigen}
H^0 & = & \phi_1^0 \cos\alpha + \phi^0_2\sin\alpha ,\,\,\, h^0  =
-\phi^0_1\sin\alpha + \phi^0_2 \cos\alpha   ,
 \eeq
 where $\alpha$ is the mixing angle.

After diagonalizing the mass matrix of the fermion fields, the
Yukawa Lagrangian becomes~\cite{David}
 \bea  L_Y&=&-\overline{U}M_UU-\overline{D}M_DD
 +\frac{i}{\upsilon}\chi^0\left(
 \overline{U}M_U\gamma_5U-\overline{D}M_D\gamma_5D\right)\nnb \\
 &+&\frac{\sqrt{2}}{\upsilon}
\chi^-\overline{D}V^\dagger_{CKM}\left[M_UR-M_DL\right]U
-\frac{\sqrt{2}}{\upsilon}\chi^+
\overline{U}V_{CKM}\left[M_DR-M_UL\right]D\nnb \\
&+&\frac{iA^0}{\sqrt{2}}\left\{\overline{U}\left[\widehat{\xi}^UR-\widehat{\xi}^{U\dag}L\right]U
+\overline{D}\left[\widehat{\xi}^{D\dag}L-\widehat{\xi}^{D}R\right]D\right\}\nnb\\
 &-&\frac{H^0}{\sqrt{2}}\overline{U}\left\{\frac{\sqrt{2}}{\upsilon} M_U
 \cos\alpha+\left[\widehat{\xi}^UR+\widehat{\xi}^{U\dag}L\right]\sin\alpha\right\}U
 -\frac{H^0}{\sqrt{2}}\overline{D}\left\{\frac{\sqrt{2}}{\upsilon} M_D
 \cos\alpha+\left[\widehat{\xi}^DR+\widehat{\xi}^{D\dag}L\right]\sin\alpha\right\}D\nnb\\
 &-&\frac{h^0}{\sqrt{2}}\overline{U}\left\{-\frac{\sqrt{2}}{\upsilon} M_U
 \sin\alpha+\left[\widehat{\xi}^UR+\widehat{\xi}^{U\dag}L\right]\cos\alpha\right\}U
-\frac{h^0}{\sqrt{2}}\overline{D}\left\{\frac{\sqrt{2}}{\upsilon}M_D
 \sin\alpha+\left[\widehat{\xi}^DR+\widehat{\xi}^{D\dag}L\right]\cos\alpha\right\}D\nnb\\
 &-&H^+\overline{U}\left[V_{CKM}\widehat{\xi}^DR-\widehat{\xi}^{U\dag}V_{CKM}L\right]D
 -H^-\overline{D}\left[\widehat{\xi}^{D\dag}V^\dagger_{CKM}L-V^\dagger_{CKM}\widehat{\xi}^UR\right]U,
 \label{lyukmass}
  \eea
where U and D are now the fermion mass eigenstates and
 \bea
\hat\eta^{U,D}&=&(V_L^{U,D})^{-1}\cdot \eta^{U,D} \cdot
V_R^{U,D}=\frac{\sqrt{2}}{v}M^{U,D}(M^{U,D}_{ij}=\delta_{ij}m_j^{U,D}),
\label{diag}\\
\hat\xi^{U,D}&=&(V_L^{U,D})^{-1}\cdot \xi^{U,D} \cdot V_R^{U,D}
\label{neutral},
 \eea
 where $V_{L,R}^{U,D}$ are
the rotation matrices acting on up- and down-type quarks, with left-
and right-chiralities respectively. Thus,
$V_{CKM}=(V_L^U)^{\dag}V_L^D$ is the usual Cabibbo-Kobayashi-Maskawa
(CKM) matrix. In general, the matrices $\hat\eta^{U,D}$ in
Eq.~(\ref{diag}) are diagonal, while the matrices $\hat\xi^{U,D}$ are
 non-diagonal which could induce scalar-mediated FCNC. From
 Eq.~({\ref{lyukmass}}), we obtain that the couplings of neutral Higgs bosons to
the fermions could generate the FCNCs.
  For the arbitrariness of definition for $\xi^{U,D}_{ij}$ couplings,
we can adopt the rotated couplings expressed $\xi^{U,D}$ in stead of
$\hat{\xi}^{U,D}$ hereafter.

In this paper, we use the Cheng-Sher ansatz~\cite{cheng} \be
\xi^{U,D}_{ij}=\lambda_{ij} \,\frac{\sqrt{2}\sqrt{m_i
m_j}}{v}\label{ans}, \ee
 which ensures that the FCNCs within the first two
generations are naturally suppressed by small fermions masses. This
ansatz suggests that LFV couplings involving the electron are
 suppressed, while LFV transitions involving muon and tau
are much less suppressed and may lead to some effects which are
promising to be tested by the future B factory experiments. In
Eq.~(\ref{ans}), the parameters $\lambda_{ij}$ are complex
 where $i,j$ are the generation indexes. In this study,
we will discuss $\tau^- \to \mu^- PP$ decays
 in the 2HDM III.

\section{\bf The decay amplitudes for $\tau^- \to \mu^- PP$ decays }
In 2HDM model III, due to the existence of FCNCs, the $\tau^- \to
\mu^- PP$ processes could occour at tree level with the neutral
Higgs bosons $H^0, h^0$ and $A^0$ mediated. The decay amplitudes
could be factorized into leptonic vertex corrections and hadronic
parts described with hadronic matrix elements, which be expressed
as: \bea \langle \mu^-PP |{\cal M} |\tau^- \rangle
&=&\frac{iG_F}{2\sqrt{2}} \cdot m_q\sqrt{m_\tau m_\mu}\cdot
\left\{\biggl[ H^{q*}\cdot \lambda_{\tau\mu} \cdot
(\bar{\mu}\tau)_{S+P}+ H^{q}\cdot \lambda^*_{\tau\mu}\cdot
(\bar{\mu}\tau)_{S-P}\biggl]\cdot<PP|(\bar{q}q)_S|0>
 \right.\nnb
\\&&\left.+\biggl[ N^{q*}\cdot \lambda_{\tau\mu}\cdot
(\bar{\mu}\tau)_{S+P}- N^{q}\cdot \lambda^*_{\tau\mu}\cdot
(\bar{\mu}\tau)_{S-P}\biggl]\cdot<PP|(\bar{q}q)_P|0> \right\},
\label{amplitude}\eea where $m_q(q=u,d,s)$ are the masses of quarks
which
 constitute the meson in the final states. $m_{\tau}$ and $m_\mu$ are the masses of lepton
 $\tau$ and $\mu$. $S\pm P$ denotes $(1\pm \gamma_5)$. The auxiliary functions
 $H^q$ and $N^q$ can be written as
\bea
 H^u&=&(A+B)\cdot \lambda^*_{uu}+(A-B) \cdot \lambda_{uu}
+F,\,\, H^{d,s}=(A-B)\cdot \lambda^*_{dd(ss)}+(A+B) \cdot
\lambda_{dd(ss)}
+F,\nnb \\
N^u&=&(A+B)\cdot \lambda^*_{uu}-(A-B) \cdot \lambda_{uu},\,\,\,
N^{d,s}=(A-B)\cdot \lambda^*_{dd(ss)}-(A+B) \cdot \lambda_{dd(ss)}
,\nnb \\
A&=&\frac{\sin^2 \alpha}{m^2_{H^0}}+ \frac{\cos^2
\alpha}{m^2_{h^0}},\,\,\,\,\,\, B=\frac{1}{m^2_{A^0}},\,\,\,\,\,\,
F=2\sin \alpha \cos\alpha (\frac{1}{m^2_{H^0}}- \frac{1}{m^2_{h^0}})
,\label{fun} \eea where $m_{H^0}$, $m_{h^0}$ and $m_{A^0}$ are the
masses of neutral Higgs bosons, $\alpha$ is the mixing angle.
$\lambda_{\tau\mu}$ is the LFV coupling parameter of neutral Higgs
bosons to lepton $\tau$ and $\mu$, and $\lambda_{uu(dd,ss)}$ are the
quark counterparts. $<PP|(\bar{q}q)_{S(P)}|0>$ are the hadronic
matrix elements. From the Eq.~(\ref{amplitude}) and (\ref{fun}), we
can see that three neutral Higgs bosons take effects to two
pseudo-scalar mesons through $(\bar{q}q)_{S\pm P}$ operators. But it
differs from the MSSM where the $\gamma$ contributions are the
dominant one and only $H^0$ and $h^0$ take effects at the case of
large $\tan \beta$~\cite{Arganda}.

Next, we will calculate the hadronic matrix elements. The ordinary
method is to parameterize the hadronic vertexes by hadron masses and
their decay constants. For the $\tau^- \to \mu^- PP$ decays, its
scale lies in the non-perturbative region and vector and scalar
resonances generally play a propelling role. However, the masses of
scalar resonances are so high that their effects are ignored.
Besides, the heavy Higgs bosons cause the hadronic final state
insensitive to resonances and their influences are less known.
Therefore, we apply $\chi PT$~\cite{Pt1,Pt2,Pt3} instead of
Resonances Chiral Theory ($R\chi T$)~\cite{Rt1,Rt2} to handle
$<PP|(\bar{q}q)_{S(P)}|0>$. Additionally, it should be noted that
the pseudo-scalar currents have contributions only to one
pseudo-scalar meson in final states. Thus, by virtue of the relevant
scalar currents $S^{i}(i=0,3,8)$ in $\chi PT$ and
 using the following currents~\cite{Arganda}:
 \bea
  -\bar{u}
u&=&\frac{1}{2}S^3+\frac{1}{2\sqrt{3}}S^8
+\frac{1}{\sqrt{6}}S^0,\nnb \\
-\bar{d}d&=&-\frac{1}{2}S^3+\frac{1}{2\sqrt{3}}S^8
+\frac{1}{\sqrt{6}}S^0,\nnb \\
-\bar{s} s&=&-\frac{1}{\sqrt{3}}S^8 +\frac{1}{\sqrt{6}}S^0, \eea we
could obtain the amplitudes: \vskip-1cm \bea {\cal M}(\tau^- \to
\mu^- PP )&=& \frac{i G_F}{2\sqrt{2}}\cdot \sqrt{m_\tau m_\mu} \cdot
[T(PP) \cdot \lambda_{\tau\mu}\cdot (\bar{\mu}\tau)_{S+P} +T^*(PP)
\cdot \lambda^*_{\tau\mu}\cdot (\bar{\mu}\tau)_{S-P}], \label{amp} \\
T(K^+K^-)&=&A\cdot[Re(\lambda_{uu})\cdot m^2_\pi+
 Re(\lambda_{ss})\cdot (2m^2_K
-m^2_\pi)]+Bi\cdot [Im(\lambda_{ss})\cdot (2m^2_K
-m^2_\pi)-Im(\lambda_{uu})\cdot m^2_\pi]\nnb \\&&+F\cdot m^2_K,
 \label{amp1}\\
T(K^0\bar{K}^0)&=&A\cdot[Re(\lambda_{dd})\cdot m^2_\pi +Re(\lambda_{ss})\cdot
(2m^2_K -m^2_\pi)]+Bi\cdot[Im(\lambda_{ss})\cdot (2m^2_K -m^2_\pi)+Im(\lambda_{dd})\cdot m^2_\pi
]\nnb \\&&+F\cdot m^2_K, \label{amp2}\\
T(\pi^+\pi^-)&=&m^2_\pi\cdot
\left\{A[Re(\lambda_{uu})+Re(\lambda_{dd})]
 +Bi\cdot[ Im(\lambda_{dd})-Im(\lambda_{uu})]+F\right\},
\label{amp3}  \\
T(\pi^0\pi^0)&=&\frac{m^2_\pi}{2\sqrt{2}}\cdot
\left\{A[Re(\lambda_{uu})-Re(\lambda_{dd})]
 -Bi\cdot[ Im(\lambda_{uu})+Im(\lambda_{dd})]\right\}
\label{amp4}, \eea where $m_\pi$ and $m_K$ are the masses of $\pi$
meson and K meson. The expressions of $A$, $B$, and $F$ can be found
in Eq.~(\ref{fun}). The $T(K^+K^-)$ and $T(K^0\bar{K}^0)$ terms are
relevant to model parameters $\lambda_{uu(ss)}$ and
$\lambda_{dd(ss)}$, respectively. The differences of
Eq.~(\ref{amp2}) from Eq.~(\ref{amp1}) are that $\lambda_{dd}$
replaces $\lambda_{uu}$ and there are an opposite sign before
$Im(\lambda_{dd})\cdot m^2_\pi$ term. While both the $T(\pi^+\pi^-)$
and $T(\pi^0\pi^0)$ terms only get involved in the parameters
$\lambda_{uu}$ and $\lambda_{dd}$.

\section{Numerical Results}

    In our calculation, the input
parameters are the Higgs masses, mixing angle $\alpha$, and couplings
 $\lambda_{qq}(q=u,d,s), \lambda_{\tau\mu}$, and their phase
 $\theta$.
We use the values of parameters in the literatures~\cite{dyb,csh}
 \vskip-1cm
\bea
m_{H^0}&=&160GeV, \,\,\,m_{h^0}=115GeV,\,\,\,
  m_{A^0}=120GeV,\,\,\,\alpha=\pi/4,\,\,\, \theta_{qq}=\pi/4,
  \eea
where the experimental constraints from $B-\bar{B}$ mixing, $b\to s
\gamma$, $\rho^0$, and $R_b$ are considered. The constraints on
$|\lambda_{uu(dd,ss)}|$ and $|\lambda_{\tau\mu}|$ from different
phenomenological
considerations~\cite{csh,Martin,Atwood1,Atwood2,Atwood3,Rodolfo03,Cotti}
are demonstrated in Table~I. The constraints on $|\lambda_{ss}|$
from $B_s-\bar{B}_s$ mixing and $b \to s \gamma $ process, are
$\gtrsim {\cal O}(1)$~\cite{csh}. Whereas for the first generation,
the FCNC couplings are suppressed, the values of $\lambda_{uu}$ and
$\lambda_{dd}$ are less than 1~\cite{Atwood3}. For parameter in
lepton sector, the $\tau \to 3\mu, \tau \to \mu \gamma, h^0\to
f\bar{f}$ decays and $(g-2)_\mu$ have constrainted
$\lambda_{\tau\mu}$ at the order of ${\cal O}(10)-{\cal
O}(10^2)$~\cite{Rodolfo03,Martin,Cotti}. Hence, for the following
values$^{[*]}$ of $|\lambda_{\tau\mu}|=150,\,\,\,
|\lambda_{uu(dd)}|\sim (0,0.9),\,\,\,|\lambda_{ss}|\sim (1,5)$, the
decay ratios are calculated.

\footnotetext[1]{The constraint on $|\lambda_{\tau\mu}|$ is found to
be $|\lambda_{\tau\mu}| \leq 2.76$ in Ref.~\cite{li1}, which is
because there the values of $|\lambda_{uu}| = 150$ and
$|\lambda_{dd}| =120$, completely different from those used in this
paper, are used.}

First, our calculations show that these branching ratios are
dependent on $\theta_{\tau\mu}$, the phase angle of
$\lambda_{\tau\mu}$. As in literatures, we assume the value of
$\theta_{\tau\mu}$ as $\pi/4$.

 \begin{table}[htb]\label{ql}
 \centering
  \begin{threeparttable}[htb]
 \caption{Constraints on the $\lambda_{ij}$ in quark and lepton
sectors.}
\begin{tabular}{c| c |c |c}
\hline\hline&Bounds and
restrictions& Process and Restriction& References \\
\hline
 Quark sector  &$|\lambda_{ss}| \gtrsim {\cal O}(1)$
 &$B_s-\bar{B}_s$ mixing,\,\, $b \to s \gamma$ &~\cite{csh}
 \\
  \cline{2-4}&$|\lambda_{uu}|,|\lambda_{dd}| \sim {\cal O}(1)$
 &$F^0-\bar{F^0}$ mixing ($F=K,B_d,D$),
 $R_b,\rho,B\to X_s \gamma $&~\cite{Atwood3}\\ \cline{1-4}
&$\lambda_{\tau\mu}\sim {\cal O}(10)-{\cal O}(10^2)$
&$(g-2)_\mu,m_{A^0}\longrightarrow \infty$& ~\cite{Rodolfo03}$^{**}$ \\
\cline{2-4} Lepton sector&$|\lambda_{\mu\mu}|=|\lambda_{\tau\tau}|=
 |\lambda_{\mu\tau}|=|\lambda_{e\mu}|=10$&
 $h^0 \to f\bar{f}$& ~\cite{Martin} \\
 \cline{2-4}
 &$\lambda_{\tau\mu} \sim {\cal O}(10^2)-{\cal O}(10^3)$&
 $\tau \to 3\mu,\tau \to \mu \gamma$&~\cite{Cotti}$^{**}$\\
\hline \hline
\end{tabular}
  \begin{tablenotes}
    \item [**] Note that the constraints in~\cite{Rodolfo03} and
\cite{Cotti} are denoted by our notation.
   \end{tablenotes}
  \end{threeparttable}
\end{table}

  The Fig.~1 shows when $|\lambda_{\tau\mu}|=150$, the functions
  of $Br(\tau^- \to \mu^- K  \bar{K})$
   versus model parameters, where (a) is for $\tau^- \to \mu^-
K^+K^-$ decay and (b) is for $\tau^- \to \mu^- K^0\bar{K}^0$ decay.
One can see from Fig.~1(a) that $Br(\tau^- \to \mu^- K^+K^-)$ raises
with the increase of $|\lambda_{ss}|$ but does not vary with
$|\lambda_{uu}|$ growing. For $\tau^- \to \mu^- K^0\bar{K}^0$ decay,
the same is as that of $\tau^- \to \mu^- K^+K^-$ decay except that
$|\lambda_{dd}|$ replaces $|\lambda_{uu}|$. In the mentioned
parameter space, their decay ratios could reach the experimental
upper limits. Thus, with fixed value of $|\lambda_{\tau\mu}|$ ,
$Br(\tau^- \to \mu^- K\bar{K})$ are insensitive to the
 variation of $|\lambda_{uu(dd)}|$ and the contributions of $|\lambda_{ss}|$
  are larger than those of
 $|\lambda_{uu(dd)}|$. As can directly be seen from the Eq. (12) and (13), these results
are expected because the squared mass of K is significantly larger
than the squared mass of pion unless the value of
$|\lambda_{uu(dd)}|$ is significantly larger than $|\lambda_{ss}|$.

   With the upper bounds of $Br(\tau^- \to \mu^- K\bar{K})$,
we take $|\lambda_{uu}|=|\lambda_{dd}|=0.5$, and draw the constraint
on $|\lambda_{\tau\mu}|$ and $|\lambda_{ss}|$. Their correlation is
shown in Fig.~2,
  where the dependences of $|\lambda_{\tau\mu}|$ on $|\lambda_{ss}|$
  are denoted by the solid line for
  $\tau^- \to \mu^- K^+K^-$ decay and the dotted line
  for $\tau^- \to \mu^- K^0\bar{K}^0$ decay,
    respectively. One could get from Fig.~2 that, under the current experimental bounds,
when $|\lambda_{\tau\mu}|$ becomes large, the $|\lambda_{ss}|$ goes
to decrease, and vice versa. For example, the LFV coupling
    $|\lambda_{\tau\mu}|=10$, the upper
limits of $Br(\tau^- \to \mu^- K^+K^-)$ and $Br(\tau^- \to \mu^-
K^0\bar{K}^0)$ on $|\lambda_{ss}|$ are at $40$ and $60$,
respectively. When $|\lambda_{\tau\mu}|=400$, the values of
$|\lambda_{ss}|$ are bounded at $1$ for $\tau^- \to \mu^- K^+K^-$
decay and $1.4$ for $\tau^- \to \mu^- K^0\bar{K}^0$ decay.
Obviously, the upper bound from $Br(\tau^- \to \mu^- K^+K^-)$ on
$|\lambda_{ss}|$ is more stringent than that from $Br(\tau^- \to
\mu^- K^0\bar{K}^0)$. And in the area under the solid curve, all the
values of $|\lambda_{\tau\mu}|$ and $|\lambda_{ss}|$ are allowed.

\begin{figure}[thbp]
 \includegraphics[scale=0.40]{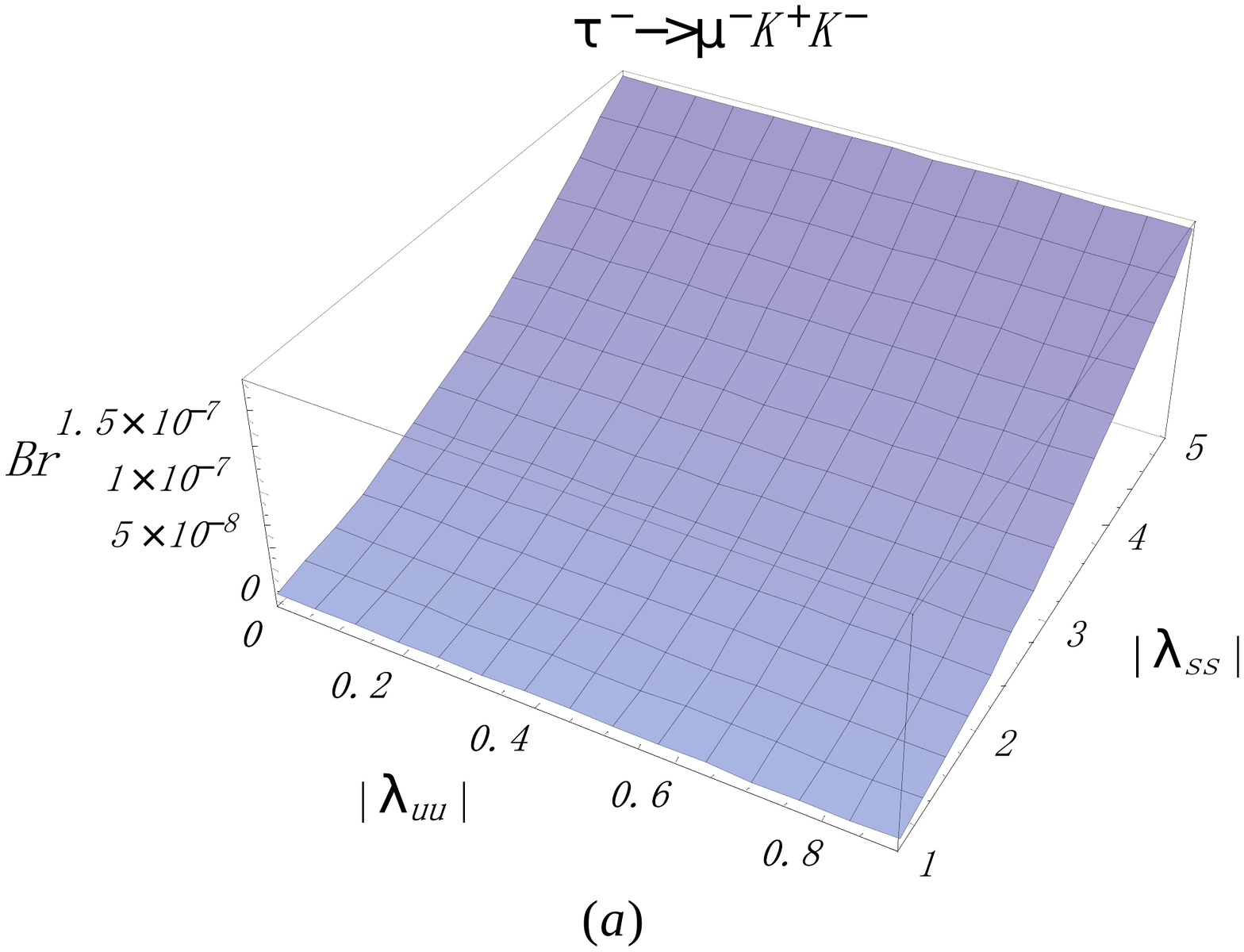}%
 \includegraphics[scale=0.4]{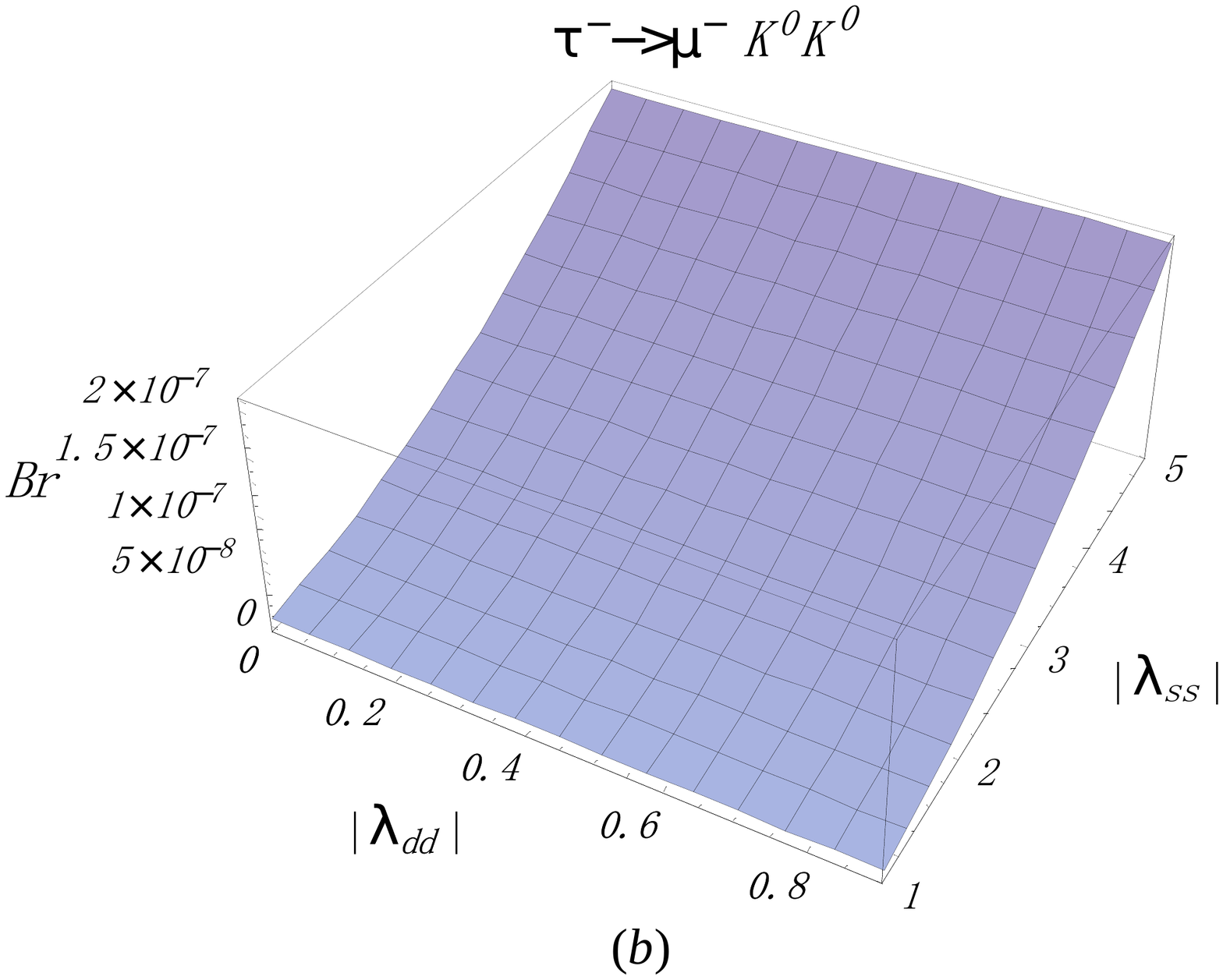}
 \vskip-6cm\caption{For $|\lambda_{\tau\mu}|=150$, the functions of branching
 ratios versus model parameters. Here, (a)
is for $\tau^- \to \mu^- K^+K^-$ decay,
 while (b) is
for $\tau^- \to \mu^- K^0\bar{K}^0$ decay. }
 \end{figure}
\vskip-0.5cm
 \begin{figure}[thbp]
 \includegraphics[scale=0.80]{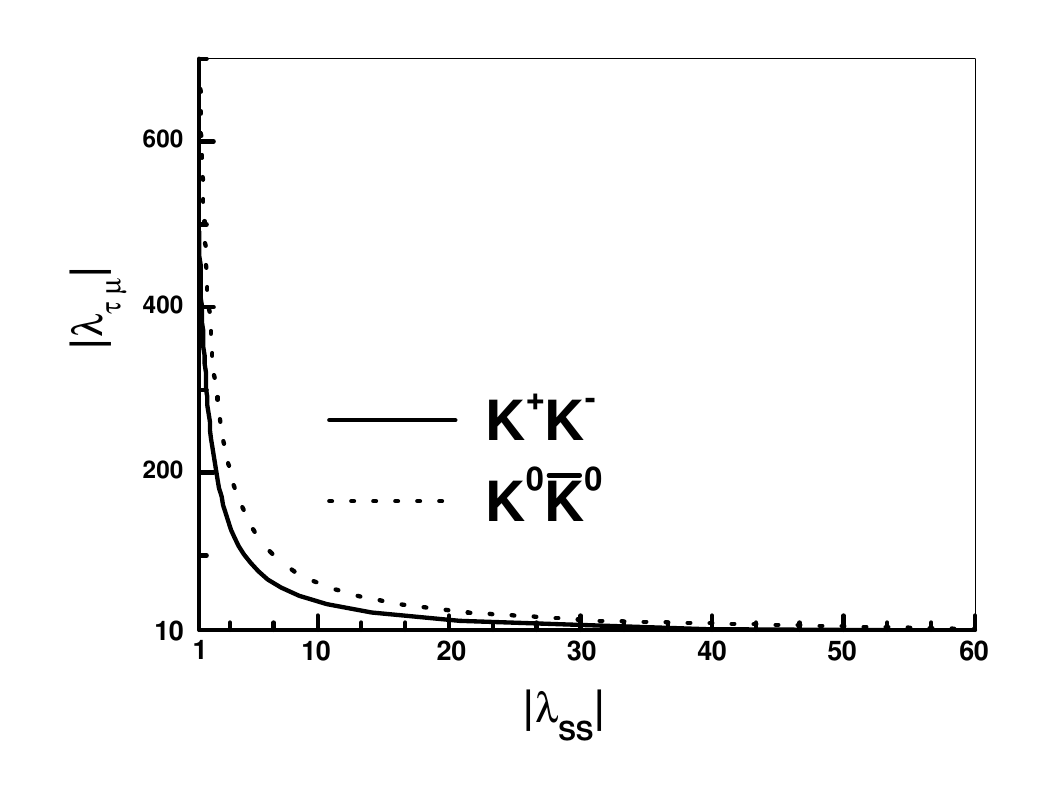}
\vskip-1cm\caption{The constraint on $|\lambda_{\tau\mu}|$ and
$|\lambda_{ss}|$ with the experimental upper bounds on $Br(\tau^-
\to \mu^- K^+K^-)$ and  $Br(\tau^- \to \mu^- K^0\bar{K}^0)$
 which read $Br(\tau^- \to \mu^-
K^+K^-)<6.8 \times 10^{-8}(CL=90\%CL)$ and
$Br(\tau^- \to \mu^- K^0\bar{K}^0)<1.6 \times
10^{-7}(CL=90\%CL)$.}\end{figure}

 The functions of $Br(\tau^- \to \mu^- \pi\pi)$ versus
$|\lambda_{uu}|$ and $|\lambda_{dd}|$ are presented in Fig.~3, where
$|\lambda_{\tau\mu}|=150$, (a) is for $\tau^- \to \mu^- \pi^+\pi^-$
decay, and (b) is for $\tau^- \to \mu^- \pi^0\pi^0$ decay. In
Fig.~3, with $|\lambda_{\tau\mu}|=150$, $Br(\tau^- \to \mu^-
\pi^+\pi^-/\pi^0\pi^0)$ move upward with the rising of
$|\lambda_{uu}|$ and $|\lambda_{dd}|$, which is different from the
case of $\tau^- \to \mu^- K \bar{K}$ decays. But their decay ratios
only reach to the order of ${\cal O}(10^{-11})$. Even when
$|\lambda_{\tau\mu}|=450$, the values of $Br(\tau^- \to \mu^-
\pi^+\pi^-/\pi^0\pi^0)$ are only improved one magnitude of order.
Apparently, in order to reach the experimental data,
$|\lambda_{\tau\mu}|$ should be enhanced to ${\cal O}(10^3)$ and
more. Furthermore, comparing to Fig.~3(b), the branching ratio of
$\tau^- \to \mu^- \pi^+\pi^-$ decay keeps a relatively rapid growth
and is more sensitive to $|\lambda_{uu}|$ and $|\lambda_{dd}|$.

\begin{figure}[thbp]
 \includegraphics[scale=0.40]{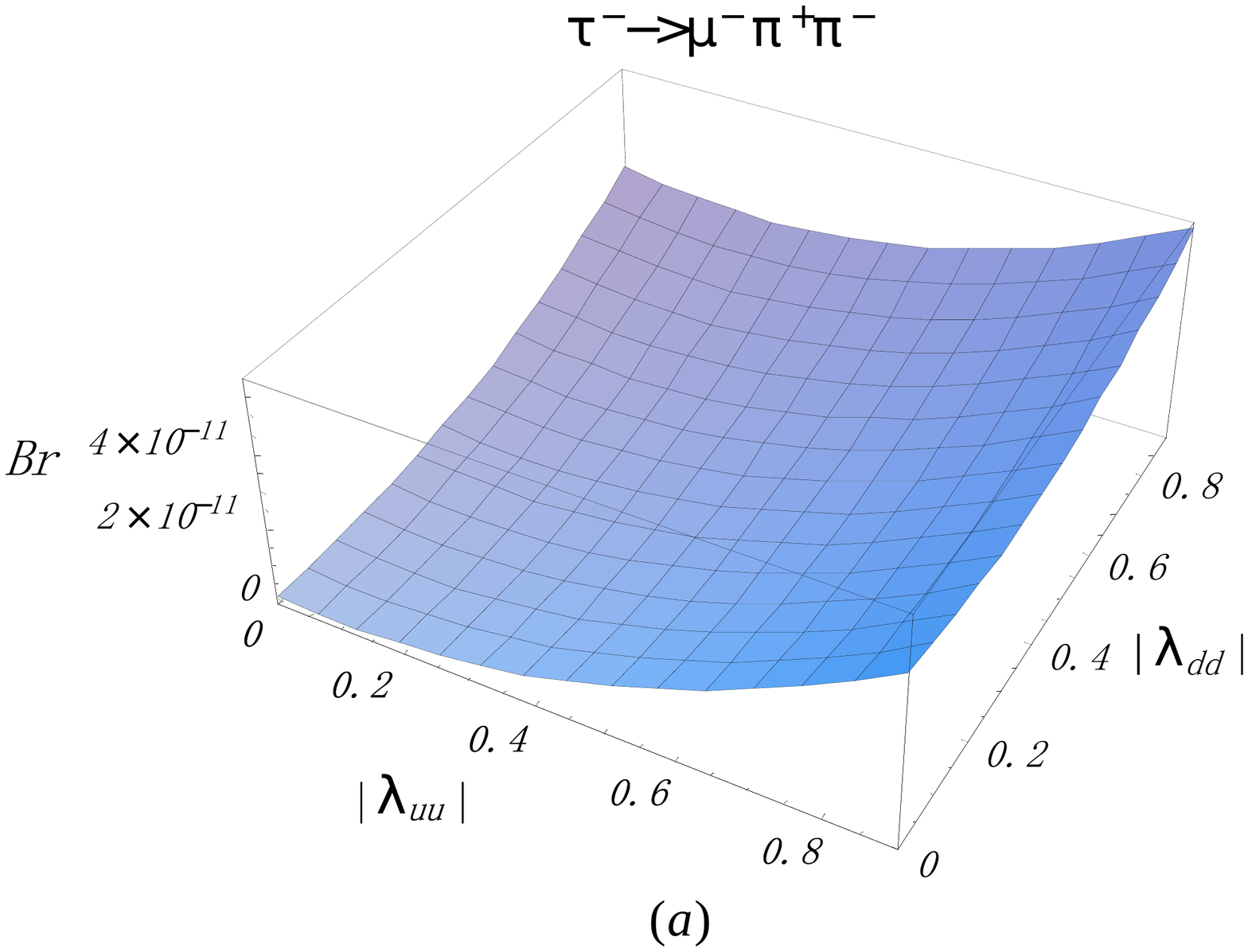}%
\includegraphics[scale=0.40]{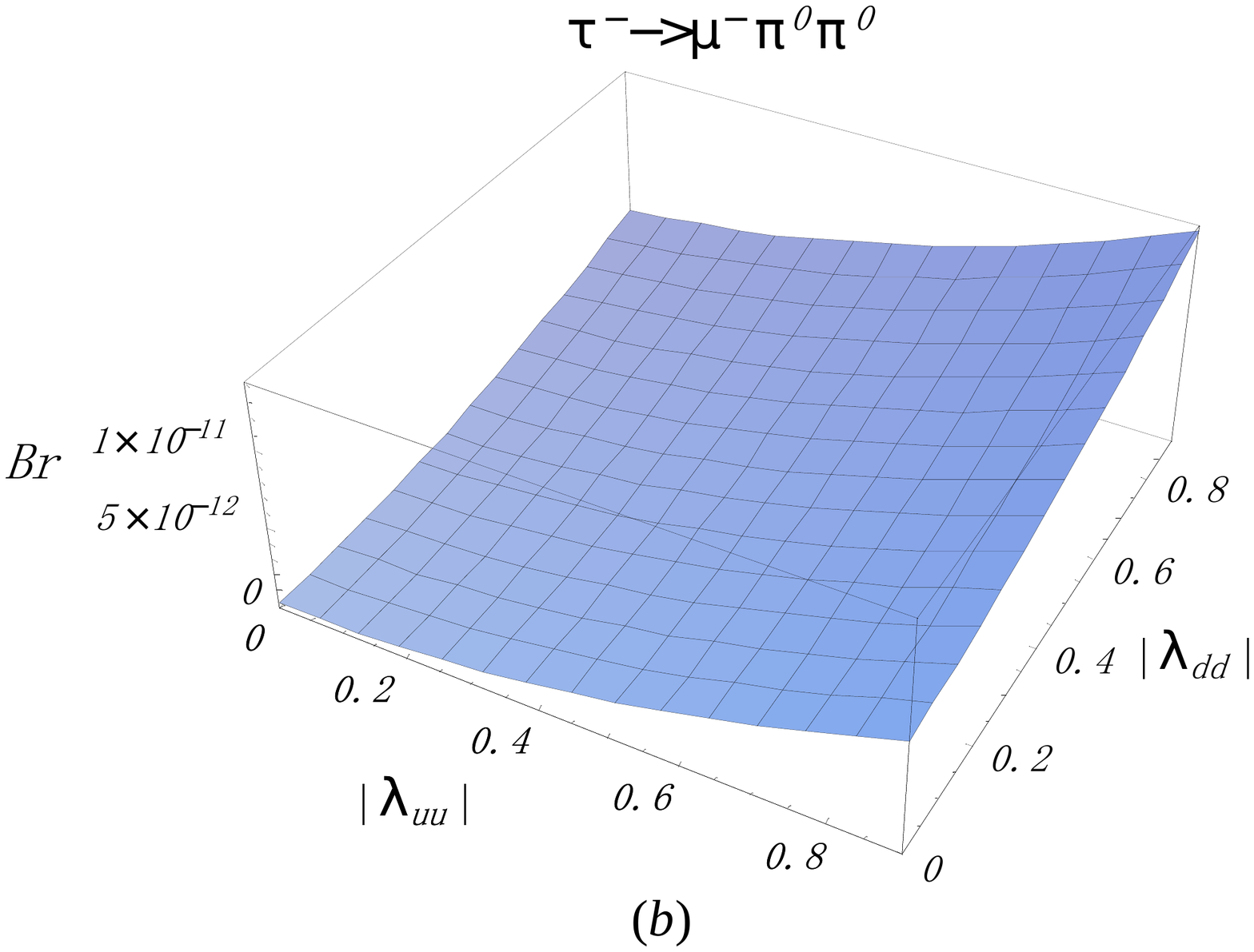}
 \vskip-6cm\caption{For $|\lambda_{\tau\mu}|=150$, the functions of branching
 ratios versus model parameters. Here, (a)
is for $\tau^- \to \mu^- \pi^+\pi^-$ decay, while (b) is for $\tau^-
\to \mu^- \pi^0\pi^0$ decay.  }
 \end{figure}

\section{\bf Conclusion}

In this paper, we have calculated the branching ratios of $\tau^-
 \to \mu^- PP(PP=K^+K^-,K^0\bar{K}^0,\pi^+\pi^-,\pi^0\pi^0)$
 decays in the 2HDM III. Because there are FCNCs
at the tree level,
 all three neutral Higgs bosons have mediated theses decays
by scalar and pseudoscalar currents. This is different from the
 MSSM where only $H^0$ and $h^0$ play roles in these processes.
Since pseudoscalar currents have no contributions to two
pseudoscalar mesons, we only considered the case of scalar currents.
  Although resonances play a role in $\tau^-\to
\mu^- PP$ processes, the massive Higgs bosons are not sensitive to
them. Consequently, we disregarded the vector resonance and used
$\chi PT$ to handle the hadron matrix elements. Our results
suggested that, the $|\lambda_{uu(dd)}|$ terms contribute really
tiny to $Br(\tau^-\rightarrow \mu^- P P$) due to the constraints on
$|\lambda_{uu(dd)}|$ from known experiments. As expected,
$Br(\tau^-\rightarrow \mu^- K^+ K^-$) could give a stringent
constraint on the parameters $|\lambda_{ss}|$ and
$|\lambda_{\tau\mu}|$. We have used its experimental upper limit to
obtain the correlation between $|\lambda_{ss}|$ and
$|\lambda_{\tau\mu}|$ and consequently the corresponding allowed
region of $|\lambda_{ss}|$ for the given region of
$|\lambda_{\tau\mu}|$. In the suitable parameter space satisfying
all known constraints, $Br(\tau^-\rightarrow \mu^- K^+ K^-$) could
be up to ${\cal O}(10^{-8})$. While for $\tau^-\rightarrow
\mu^-\pi\pi$ decays, the branching ratios are too small to be
observed. It should be noted that the latest experimental values are
based on 671 $fb^{-1}$ data. We expect that the run of Super-B
factory provides more and more data of LFV $\tau$ decays and
consequently to obtain the more stringent constraints on the model
parameters.

\begin{acknowledgments}

 The author Wen-Jun Li is indebted to Prof. Jorge Portoles, Prof. Chao-Shang Huang and Prof.
 Tian-Jun Li
for valuable discussion and wishes to thank Institute of Theoretical
Physics(ITP), Chinese Academy of Sciences and Kavli Institute for
Theoretical Physics China at the Chinese Academy of Sciences (KITPC)
for their warm accommodation. The work is supported by National
Science Foundation under contract No.10547110, No.10975171,
No.10947020, Henan Educational Committee Innovative Research Team
Foundation under contract No.2010IRTSTHN002, He¡¯nan Educational
Committee Foundation under contract No.2007140007.

\end{acknowledgments}

\end{document}